\documentclass[12pt]{article}
\usepackage{a4}
\usepackage{latexsym} 
\usepackage{amssymb}  
\usepackage{amsmath} 
\usepackage{graphicx}
\usepackage{epsfig,wrapfig}
\usepackage[ usenames]{color}
\usepackage[rflt]{floatflt} 
\usepackage{cite}
\begin{document}

\newcommand{\talk}[3]
{\noindent{#1}\\ \mbox{}\ \ \ {\it #2} \dotfill {\pageref{#3}}\\[1.8mm]}
\newcommand{\stalk}[3]
{{#1} & {\it #2} & {\pageref{#3}}\\}
\newcommand{\snotalk}[3]
{{#1} & {\it #2} & {{#3}n.r.}\\}
\newcommand{\notalk}[3]
{\noindent{#1}\\ \mbox{}\ \ \ {\it #2} \hfill {{#3}n.r.}\\[1.8mm]}
\newcounter{zyxabstract}     
\newcounter{zyxrefers}        

\newcommand{\newabstract}
{\clearpage\stepcounter{zyxabstract}\setcounter{equation}{0}
\setcounter{footnote}{0}\setcounter{figure}{0}\setcounter{table}{0}}

\newcommand{\talkin}[2]{\newabstract\label{#2}\input{#1}}                

\newcommand{\rlabel}[1]{\label{zyx\arabic{zyxabstract}#1}}
\newcommand{\rref}[1]{\ref{zyx\arabic{zyxabstract}#1}}

\renewenvironment{thebibliography}[1] 
{\section*{References}\setcounter{zyxrefers}{0}
\begin{list}{ [\arabic{zyxrefers}]}{\usecounter{zyxrefers}}}
{\end{list}}
\newenvironment{thebibliographynotitle}[1] 
{\setcounter{zyxrefers}{0}
\begin{list}{ [\arabic{zyxrefers}]}
{\usecounter{zyxrefers}\setlength{\itemsep}{-2mm}}}
{\end{list}}

\begin{center}
{\begin{flushleft}
{\tiny DESY 11-050} \hfill {\tiny LPN11-16} \\
\vspace*{-2mm} \noindent
{\tiny DO 11/05} \hfill {\tiny } 
\end{flushleft}
\large\bf \boldmath $\alpha_s(M_Z^2)$ in NNLO Analyses of Deep-Inelastic 
World Data}\\[0.5cm]
{S. Alekhin$^{1,2}$, J. Bl\"umlein$^1$, H. B\"ottcher$^1$, and S.-O. 
Moch$^1$}\\[0.3cm]
$^1$ Deutsches Elektronen-Synchrotron, DESY, Platanenallee 6, D-15738 Zeuthen, 
Germany \\
$^2$ Institute for High Energy Physics, 142281 Protvino, Moscow
Region, Russia
\end{center}

\noindent
The present world data of deep-inelastic scattering (DIS) 
reached 
a precision 
which allows the 
measurement of $\alpha_s(M_Z^2)$ from their scaling violations with an error of 
$\delta\alpha_s(M_Z^2) \simeq 1\%$. This requires at least NNLO analyses, since
NLO fits exhibit scale uncertainties of $\Delta_{r,f} \alpha_s(M_Z^2) \sim 0.0050$.
The NNLO values for $\alpha_s$ obtained are
summarized in the following Table.
{\small
{\renewcommand{\arraystretch}{1}{
\begin{center}
\begin{tabular}{|l|l|l|}
\hline
\multicolumn{1}{|c|}{ } &
\multicolumn{1}{c|}{$\alpha_s({M_Z^2})$} &
\multicolumn{1}{c|}{  } \\
\hline
BBG      & $0.1134 {\tiny{\begin{array}{c} +0.0019 \\
           -0.0021 \end{array}}}$
         & {\rm valence~analysis, NNLO}  \cite{BBG}           
\\
GRS      & $0.112 $ & {\rm valence~analysis, NNLO}  \cite{Gluck:2006yz}           
\\
ABKM           & $0.1135 \pm 0.0014$ & {\rm HQ:~FFNS~$N_f=3$} \cite{Alekhin:2009ni}             
\\
ABKM           & $0.1129 \pm 0.0014$ & {\rm HQ:~BSMN-approach} 
\cite{Alekhin:2009ni}             
\\
JR       & $0.1124 \pm 0.0020$ & {\rm
dynamical~approach} \cite{JimenezDelgado:2008hf}   
\\
JR       & $0.1158 \pm 0.0035$ & {\rm
standard~fit}  \cite{JimenezDelgado:2008hf}    
\\
MSTW & $0.1171\pm 0.0014$ &  \cite{Martin:2009bu}     \\
ABM            & $0.1147\pm 0.0012$ &   FFNS, incl. combined H1/ZEUS data   
\cite{Alekhin:2010iu}
\\
\hline
Gehrmann et al.& {{$0.1153 \pm 0.0017 \pm 0.0023$}} & {\rm
$e^+e^-$~thrust}~\cite{Gehrmann:2009eh}
\\
Abbate et al.& {{$0.1135 \pm 0.0011 \pm 0.0006$}} & {\rm
$e^+e^-$~thrust}~\cite{Abbate:2010xh}
\\
\hline
BBG & {{$
0.1141 {\tiny{\begin{array}{c} +0.0020 \\
-0.0022 \end{array}}}$}}
& {\rm valence~analysis, N$^3$LO}  \cite{BBG}            \\
\hline
{world average} & {$
0.1184 \pm 0.0007$  } & \cite{Bethke:2009jm}
\\
\hline
\end{tabular}
\end{center}
\renewcommand{\arraystretch}{1}   
}}
}
\noindent
NNLO non-singlet data analyses have been performed in 
\cite{BBG,Gluck:2006yz}. 
The analysis of Ref.~\cite{BBG} is based on an experimental combination of flavor 
non-singlet data referring to $F_2^{p,d}(x,Q^2)$ for $x < 0.35$ and using the 
respective valence approximations for $x > 0.35$. The $\overline{d} - \overline{u}$ 
distributions and the $O(\alpha_s^2)$ heavy flavor corrections were accounted for. 
At low $Q^2$ 
and at large $x$ also at low $W^2$ higher twist corrections have to be taken into 
account \cite{HT}. The corresponding region was cut out in \cite{BBG} performing 
the fits for the leading twist terms only. The analysis 
could be extended to N$^3$LO effectively due to the dominance of the Wilson 
coefficient in this order \cite{Vermaseren:2005qc} if compared to the anomalous 
dimension, 
cf.~\cite{BBG,Baikov:2006ai}. 
This analysis led to an increase of
$\alpha_s(M_Z^2)$ by $+0.0007$ if compared to the NNLO value.

A combined singlet and non-singlet NNLO analysis based on the DIS world data, 
including the Drell-Yan and di-muon data, needed for a correct description of 
the sea-quark densities, was performed in \cite{Alekhin:2009ni}. In the fixed 
flavor number scheme (FFNS) the value of $\alpha_s(M_Z^2)$ is the same as in 
the non-singlet case \cite{BBG}. The comparison between the FFNS and the BMSN 
scheme \cite{BMSN} 
for the description of the heavy flavor contributions induces a 
systematic uncertainty $\Delta \alpha_s(M_Z^2) = 0.0006$. The NNLO analyses of 
Ref.~\cite{JimenezDelgado:2008hf} are statistically compatible with the results
of  
\cite{BBG,Gluck:2006yz,Alekhin:2009ni}, 
while those of \cite{Martin:2009bu} yield a 
higher value.

In Ref.~\cite{Alekhin:2010iu} the combined H1 and ZEUS data were accounted for 
in a NNLO analysis for the first time, which led to a shift of $+0.0012$. However,
running quark mass effects \cite{Alekhin:2010sv}
and the account of recent $F_L$ data reduce this value 
again to 
the NNLO value given in \cite{Alekhin:2009ni}. We mention that other recent NNLO
analyses of precision data, as the measurement of $\alpha_s(M_Z^2)$ using thrust 
in high energy
$e^+e^-$ annihilation data  \cite{Gehrmann:2009eh,Abbate:2010xh}, result in lower 
values than the 2009 world average \cite{Bethke:2009jm} based on NLO, NNLO and 
N$^3$LO results. The sensitivity of the fits to a precise description of the 
longitudinal structure function $F_L$ has been demonstrated in 
\cite{Alekhin:2011ey} recently, in the case of the NMC data. Inconsistent 
descriptions of $F_L$ induce a high value of $\alpha_s$ of $\sim 0.1170$ to be 
compared with that obtained in \cite{Martin:2009bu}. 
It is observed that the values of 
{\small \renewcommand{\arraystretch}{1.0}
\begin{center}
\begin{tabular}{|r|r|r|l|}
\hline
$\alpha_s(M_Z^2)$
& with $\sigma_{\rm NMC}$
& with $F_2^{\rm NMC}$
& difference \\
\hline
NNLO         &  {0.1135(14)} & {0.1170(15)} & $+0.0035 \simeq 2.3 \sigma$
\\
NNLO +{$ F_L$} $O(\alpha_s^3)$
             &  0.1122(14) & 0.1171(14) & $+0.0050 \simeq 3.6 \sigma$
\\
\hline
\end{tabular}
\end{center}
\renewcommand{\arraystretch}{1.0}}

\vspace*{2mm}\noindent
$\alpha_s$ found
in NLO fits are
systematically higher than those in NNLO analyses. $\alpha_s$ measurements based on 
jet 
data can be performed presently at NLO only. Here typical values 
obtained are
$\alpha_s(M_Z^2) = 0.1156 \tiny \begin{array}{c} +0.0041 \\ - 0.0034
\end{array}$ \cite{Frederix:2010ne}, 
$\alpha_s(M_Z^2) = 0.1161 \tiny \begin{array}{c} +0.0041 \\ - 0.0048
\end{array}$ \cite{Abazov:2009nc} in recent examples. The precise knowledge of
$\alpha_s(M_Z^2)$ is of instrumental importance for the correct prediction of the 
Higgs 
boson cross section at Tevatron and the LHC \cite{Alekhin:2010dd}.

\vspace*{-4mm}
{\small

}

%

\end{document}